\documentclass[universe,article,submit,pdftex,moreauthors]{Definitions/mdpi}

\firstpage{1}
\pubvolume{1}
\issuenum{1}
\articlenumber{0}
\pubyear{2024}
\copyrightyear{2024}
\datereceived{ }
\dateaccepted{ }
\datepublished{ }
\hreflink{https://doi.org/} 

\usepackage{dcolumn}                 
\usepackage{bm}                      

\def\be{\begin{equation}} \def\ee{\end{equation}}
\def\bal#1\eal{\begin{align}#1\end{align}}
\def\bse#1\ese{\begin{subequations}#1\end{subequations}}
\def\kv{\bm{k}} \def\rv{\bm{r}} \def\Rv{\bm{R}}
\def\sv{\bm{\sigma}}
\def\eps{\varepsilon}
\def\epot{E_\text{cor}}
\def\epspot{\eps_\text{cor}}
\def\sat{\text{sat}}

\def\fm3{\,\text{fm}^{-3}}
\def\tausm{\tau_\text{SM}}
\def\taunm{\tau_\text{NM}}

\long\def\OFF#1{}

\Title{
Nuclear Matter Equation of State
in the Brueckner-Hartree-Fock Approach
and Standard Skyrme Energy-Density Functionals
}

\TitleCitation{Nuclear Matter Equation of State} 

\Author{
Isaac Vida\~na $^{1,\ddagger}$, 
J\'er\^ome Margueron $^{2,3,*,\dagger,\ddagger}$ and 
Hans-Josef Schulze $^{1,\ddagger}$ 
}

\AuthorNames{Isaac Vida\~na, J\'er\^ome Margueron and Hans-Josef Schulze}

\AuthorCitation{Vida\~na, I.; Margueron, J.; Schulze, H.-J.}

\address{
$^1$\quad Istituto Nazionale di Fisica Nucleare, Sezione di Catania, Dipartimento di Fisica e Astronomia ``Ettore Majorana",
Universit\'a di Catania, Via Santa Sofa 64, I-95123 Catania, Italy\\
$^2$\quad Institut de Physique Nucl\'eaire, Universit\'e Lyon 1 and CNRS/IN2P3,
4 rue Enrico Fermi, 69622 Villeurbanne, France\\
$^3$\quad International Research Laboratory on Nuclear Physics and Astrophysics,
Michigan State University and CNRS, East Lansing, MI 48824, USA
}

\corres{Correspondence: jerome.margueron@cnrs.fr}
\secondnote{These authors contributed equally to this work.}

\abstract{
The equation of state of asymmetric nuclear matter
as well as the neutron and proton effective masses
and their partial-wave and spin-isospin decomposition
are analyzed within the Brueckner--Hartree--Fock approach.
Theoretical uncertainties for all these quantities are estimated
by using several phase-shift-equivalent nucleon-nucleon forces
together with two types of three-nucleon forces,
phenomenological and microscopic.
It is shown that the choice of the three-nucleon force plays an important role
above saturation density,
leading to different density dependencies of the energy per particle.
These results are compared to the standard form
of the Skyrme energy-density functional
and we find that it is not possible to reproduce the BHF predictions in the
$(S,T)$ channels in symmetric and neutron matter above saturation density,
already at the level of the two-body interaction,
and even more including the three-body interaction.
}


\begin{document}

\section{Introduction}
\label{sec1}

A new era of multi-messenger astronomy has begun with the first detection
of gravitational wave signals,
also known as GW170817 or AT2017gfo for the electromagnetic emission
originating from the kilonova formed by the binary neutron star merger
\cite{TheLIGOScientific:2017,LIGOScientific:2017a}.
This new astronomy emphasizes the crucial role of neutron stars (NSs),
which are prototypes for the most extreme phases of matter where the strong,
the weak, the gravitational and the electromagnetic interactions can be studied
in regimes that cannot be explored on Earth.
In this paper,
we are analyzing the properties of the dense-matter equation of state (EOS)
from microscopic calculations with a focus on the strong three-body interaction
and we discuss the link with empirical energy density functionals (EDFs).

In recent years,
the determination of nuclear EDFs incorporates more and more information
deduced from microscopic calculations of nuclear matter.
Usually,
one specific or a very limited number of microscopic calculations
are considered in the fitting protocoles.
However, the impact of the choices of
the bare nuclear interaction,
the regularization scheme,
the experimental error in the phase shift data,
and of the many-body framework is important,
and has been cautiously addressed in the chiral effective field theory
($\chi$-EFT) \cite{chift}
in order to
theoretically predict the nuclear EOS including error estimates.
There is however an upper limit on density
(presently estimated to be (1--2)$n_\sat$
with $n_\sat\approx 0.155\pm0.005\fm3$ \cite{Margueron2018a}),
where the chiral EFT breaks down
and also the error estimates become meaningless
\cite{Hebeler:2013nza,Tews:2018kmu,Drischler:2019wtt}.

At increasing density
-- and already around $n_\sat$ --
also the role of nuclear three-body forces (3BF) in nuclear matter
cannot be ignored.
For traditional meson-exchange potential models,
the compatible 3BF correction at saturation density
is of limited magnitude, a few MeV,
which represents (15--20)\% of the total energy,
in order to reproduce the saturation of nuclear matter.
With the advent of regularized soft-core nuclear interactions,
such as $V_\text{low-k}$ \cite{vlowk}
as well as chiral effective field theory ($\chi$-EFT) \cite{chift},
a new paradigm has emerged.
In short, any modification of the two-body interaction,
by applying a unitary transformation, generates many-body interactions.
The 3BF cannot be viewed as independent of the two-body force (2BF),
but instead, it contains a part of the repulsive hard-core interaction.
Regularized soft-core interactions need stronger many-body interactions
to saturate, compared to the hard-core ones.

Building bridges between EDFs for the in-medium nuclear interaction
and microscopic approaches of the nuclear many-body problem is, therefore,
fundamental for a consistent understanding of nuclei,
neutron stars, core-collapse supernovae, and kilonovae.
Microscopic approaches, on one hand, have the advantage of being based on
realistic 2BFs
that reproduce with high precision
the scattering phase shifts and the deuteron properties,
and include the isospin asymmetry dependence in a natural way.
However, the direct implementation of microscopic approaches in finite nuclei
is not an easy task.
A realistic 2BF cannot be used directly in nuclei.
Its hard-core repulsion requires a resummation as in uniform matter
and 3BFs are required to describe both the energy
and charge radii of finite nuclei.
EDF models, on the other hand, are usually based on effective
density-dependent interactions with parameters frequently fitted
to reproduce global properties of nuclei and properties
of symmetric and neutron matter
\cite{Chabanat1997,Goriely2010,Fantina2013}.
These effective interactions capture the essence of the
in-medium nuclear interaction as well as the effect of the complex correlations.
Their predictions at high densities and isospin asymmetries, however,
should be taken with care and discussed in view of the considered constraints.
Combining phenomenological models with microscopic approaches would, therefore,
help in setting up a nuclear model based on realistic 2BFs and 3BFs
that can simultaneously describe the properties of nuclei and those of
infinite nuclear matter in a large range of densities and isospin asymmetries.

Two major sources of uncertainty arise, however,
when trying to build links between EDFs and microscopic approaches.
The first is that there is a fairly large number of realistic 2BFs
that reproduce the scattering data and the deuteron properties
with equivalent high accuracy.
Predictions employing these interactions shall then be compared
to estimate the uncertainty due to the nuclear interaction
(leading to an estimate of systematic uncertainties).
The second source of uncertainties is due to the different
methods employed to solve the nuclear many-body problem
(which also contributes to systematical uncertainties).
A critical comparison of various microscopic approaches using the same 2BF
has been performed in Ref.~\cite{baldo12}.
The aim of this work is to find the sources of discrepancies,
and ultimately to determine a ``systematic uncertainty"
associated with the different microscopic approaches of the nuclear-matter EOS.
The approaches considered were the Brueckner--Hartree--Fock (BHF) \cite{BHF},
the Brueckner--Bethe--Goldstone (BBG) expansion up to third order \cite{BBG},
the self-consistent Green's function (SCGF) \cite{SCGF},
the auxiliary field diffusion Monte Carlo (AFDMC) \cite{AFDMC},
the Green's function Monte Carlo (GFMC) \cite{GFMC},
and the Fermi hypernetted chain (FHNC) \cite{FHNC}.
The properties of pure neutron matter (NM)
and symmetric nuclear matter (SM) were computed
with simplified versions \cite{SIMARG}
of the widely used Argonne Av18 potential \cite{v18},
and a careful comparison of the results obtained with the different approaches
was performed.
The results of this work confirmed that the tensor and spin-orbit components
of the 2BF and their in-medium treatment are responsible
for most of the observed discrepancies among these approaches \cite{baldo12}.
A very similar study has been performed recently in
Refs.~\cite{piarulli20,lovato22}.

The EDF models adjusted to microscopic approaches shall incorporate
these sources of uncertainties in the microscopic predictions for dense matter.
Several authors have already determined EDF models from microscopic approaches
(see, e.g., Refs.~\cite{Baldo2004,bcpm,Cao2006,Lesinski2006,Goriely2010,Gambacurta2011,Davesne2015}).
The authors of Ref.~\cite{Baldo2004}, for instance,
constructed a Skyrme-type EDF based on BHF results for infinite matter
and applied it to a set of isospin-symmetric and asymmetric nuclei
through the mass table.
Their results provided a microscopic basis for a link between
nuclear surface behavior and the NM EOS,
as previously observed with phenomenological effective forces.
The same authors have also developed the so-called
Barcelona-Catania-Paris-Madrid (BCPM) EDF \cite{bcpm}
obtained from BHF calculations of nuclear matter
within an approximation inspired by the Kohn-Sham formulation
of density functional theory \cite{KS}.
The BCPM EDF is built with a bulk part obtained directly from BHF results
for NM and SM via a local-density approximation,
and is supplemented by a phenomenological surface part
together with the Coulomb, spin-orbit, and pairing contributions.
The functional, with four adjustable parameters,
is able to describe the ground-state properties of finite nuclei
with an accuracy comparable to that of Skyrme and Gogny forces.

Later on, the so-called LNS Skyrme force was built in Ref.~\cite{Cao2006}
based also on BHF calculations of infinite nuclear matter
with consistent 2BFs and 3BFs.
It was further refined in Ref.~\cite{Gambacurta2011}
to reproduce experimental binding energies and charge radii
of some selected nuclei.
A similar fitting procedure was also used in Ref.~\cite{Goriely2010},
where various Skyrme forces were adjusted to the $(S,T)$ decomposition
of the results of two different BHF calculations \cite{Zhou2004,Li2008}
\footnote{
Note that in Fig.7 of Ref.~\cite{Goriely2010} the BHF predictions for the $(S,T)$ channels
$(0,1)$ and $(1,1)$ in SM from catania2~\cite{Zhou2004} results have been exchanged.}.
The comparison of the two microscopic calculations provided a kind of
theoretical uncertainty based on the 2BFs and 3BFs employed,
and suggests that the associated differences should be carefully analyzed.
In Ref.~\cite{Davesne2015} the tensor contribution
to the $(S,T)$ channels and different partial waves was analyzed.
However, similarly to previous works,
the uncertainty of the microscopic calculations
was not included in the determination of the EDF.
Finally let us mention the analysis of
Quantum Monte Carlo (QMC) calculations
of spin-polarized neutron matter,
which have been compared to predictions of several Skyrme EDFs
\cite{Roggero2015}.
Such analysis allowed to probe the time-odd part of the Skyrme EDFs
independently of the time-even one and was found to be very constraining.
Also here an estimate of the uncertainty of the microscopic calculation
would be very valuable,
in particular as the density of the medium increases.

In the present paper,
we aim to provide an estimation of the uncertainty of the nuclear EOS
predicted by the BHF approach up to about 0.4~fm$^{-3}$ $\gtrsim 2n_\sat$
based on several nuclear two- and three-body interactions.
Special attention is paid to the impact of the latter ones
for which we consider two different types:
phenomenological and microscopical 3BF.
We then determine the corresponding parameter ranges
of Skyrme-type nuclear energy density functionals, and more importantly,
we show that the BHF calculations
in the different $(S,T)$ channels in SM and NM
cannot be reproduced by the Skyrme EDF above $n_\sat$,
already from the 2BF results.

The paper is organized as follows.
In section~\ref{sec2} a brief review of the BHF approach is presented.
The $(S,T)$ and partial-wave decomposition of the BHF correlation energy
per particle for SM and NM is analysed in section~\ref{sec3}.
In section \ref{sec:skyrme} this decomposition is employed
to constrain the parameters of Skyrme-type forces.
A summary and the main conclusions are given in section~\ref{sec5}.

\section{BHF approach of nuclear matter}
\label{sec2}

The BHF approach is the lowest-order realization of the BBG many-body theory
of nuclear matter \cite{bbgt}.
In this theory, the ground-state energy of nuclear matter is evaluated
in terms of the so-called hole-line expansion,
where the perturbative diagrams are grouped according to the number of
independent hole-lines.
The expansion is derived by means of the
in-medium two-body scattering matrix $G$,
which describes the effective interaction between two nucleons
in the presence of a surrounding medium.
The $G$ matrix is obtained by solving the well-known Bethe-Goldstone equation
\be
 \langle i'j' |G| ij\rangle = \langle i'j' |V| ij\rangle
 + \frac{1}{\Omega}\sum_{ml} \langle i'j' |V| ml\rangle
 \frac{Q_{ml}}{\eps_i+\eps_j -\eps_m-\eps_l+i\eta}
 \langle ml |G| ij \rangle \:,
\label{e:bg}
\ee
where the multi-indices $i,j,m,l,i',j'$ indicate all the quantum numbers
(momentum $k$, spin $\sigma$ and isospin $\tau$ projections)
characterizing the two nucleons in the initial, intermediate, and final states,
$V$ denotes the bare $NN$ interaction,
$\Omega$ is the (large) volume enclosing the system,
$Q_{ml} = (1-f_m)(1-f_l)$ with
$f_i = \theta(k_F^{(\tau_i)}-|\kv_i|)$
is the Pauli operator taking into account the effect of the exclusion principle
on the scattered holes,
and $k_F^{(\tau_i)}=(3\pi^2 n_{\tau_i})^{1/3}$ is the Fermi momentum
for particle $i$ with density $n_{\tau_i}$
defined in the ground state.
The single-particle (s.p.) energy of a nucleon
with momentum $k_i$ and isospin $\tau_i$ is
\be
 \eps_i\equiv \eps_{\tau_i}(k_i) =
 \frac{k^2_i}{2m_{\tau_i}} + \mbox{Re}U_{\tau_i}(k_i) \:,
\label{e:spe}
\ee
being $U_{\tau_i}(k_i)$ the mean field felt by the nucleon $i$
due to its interaction with the other nucleons of the medium.
In the BHF approximation it is calculated through the ``on-shell'' $G$-matrix,
\be
 U_{\tau_i}(k_i) = \frac{1}{\Omega} \sum_j f_j \langle ij |G| ij\rangle_a \:,
\label{e:sp}
\ee
where the matrix elements are meant to be properly antisymmetrized.
Note that Eqs.~(\ref{e:bg},\ref{e:spe},\ref{e:sp})
should be solved self-consistently.
The momentum dependence of the s.p.~energy $\eps_{\tau_i}$
can be characterized in terms of the effective mass \cite{Jekeunne1976},
\be
 \frac{m^*_{\tau_i}(k_i)}{m} =
 \frac{k_i}{m}\Big(\frac{d\eps_{\tau_i}}{dk_i} \Big)^{-1} \:.
\label{e:effm1}
\ee

We consider in the following isospin-asymmetric nuclear matter
and for clarity we shall specify the isospin index $\tau$.
However, since we will discuss only spin-symmetric matter,
the spin up and down potentials are equal and we do not specify the spin index
in the following.
For simplicity we consider the same bare mass for neutrons and protons
$m_\tau=(m_n+m_p)/2 \equiv m$.

In the BHF approach, the total energy per particle of nuclear matter
is given by the sum of only two-hole-line diagrams
that include the effect of two-body correlations
through the $G$ matrix,
\be
 E_\text{tot} = \frac1n\sum_i f_i \frac{k^2_i}{2m}
 + \frac1{2n\Omega} \sum_{i,j} f_i f_j
 \mbox{Re}\langle ij |G| ij \rangle_a =
 \frac1n \sum_i f_i \frac{k^2_i}{2m} + \epot \:,
\label{e:ebhf1}
\ee
where the first term on the right hand side is the energy of a free Fermi gas
and the second one is the so-called BHF correlation energy, $\epot$.
We note here that the latter is usually referred to as BHF potential energy.

It has been shown in Ref.~\cite{BBG} for the $V_{18}$,
and recently confirmed in Ref.~\cite{thl} for several modern 2BFs,
that the contribution to the total energy from three-hole-line diagrams
(which account for the effect of three-body correlations)
is minimized when the so-called continuous prescription \cite{Jekeunne1976}
is adopted for the BHF s.p.~energy \eqref{e:spe}.
This is a strong indication of the convergence of the hole-line expansion
around saturation density and above.
We adopt this prescription in our calculations and limit the exploration
to densities less than 0.4$\fm3$,
where the hole-line expansion parameter $\kappa=(c/d)^3$
(with $c$ the interaction range and
$d$ the average distance between two nucleons)
is still sufficiently small \cite{thl,kappa}.

The present BHF calculations are carried out using a set of several
phase-shift-equivalent $NN$ potentials, namely,
Av18 \cite{av18},
NSC97a-f \cite{nnsc97},
a non-relativistic version of the Bonn B potential (Bonn) \cite{bonn}
and the charge-dependent Bonn potential (CD-Bonn) \cite{cdbonn}.
The predictions for the nuclear EOS based on these 2BFs
are grouped together under the name 2BF.
The group 2+3BFph rassemble the predictions of the same 2BF interactions
supplemented with a 3BF of Urbana type~\cite{urbana}
consisting of the sum of the attractive two-pion-exchange Fujita--Miyazawa force
with excitation of an intermediate $\Delta$-resonance \cite{fujita}
plus a phenomenological repulsive term.
The group 2+3BFmic is a set of microscopic potentials \cite{3bf2}
based on the Av18, Bonn~B, or Nijmegen93 \cite{nijm93} potentials,
which are also used to calculate consistently a three-body potential
including the virtual excitation of $\Delta$(1232) and $N^*$(1440) resonances
and anti-nucleons.
In both cases, for the use in the BHF approach,
these 3BF have been reduced to an effective two-nucleon density-dependent force
by averaging over the coordinates of the third nucleon \cite{3bf}.
The interested reader is referred to Refs.~\cite{3bf2,3bf3,3bf4,domenico15}
for an extensive analysis of the use and effects of 3BF in nuclear matter.

\section{Results of Brueckner--Hartree--Fock microscopic calculations}
\label{resbhf}
\label{sec3}

We discuss in this section the results of the BHF calculations of nuclear matter.

\begin{table}[t]
\caption{
Partial-wave contributions to the different $(S,T)$ channels
using the spectroscopic notation $^{2S+1}L_J$.}
\centering
\begin{tabular}{lcclc}
Name & $(S,T)$ & L & Partial waves &
Legend in Figs.~\ref{f:epwL} and \ref{f:epwnL} \\
\hline
Singlet Odd  & (0,0) & 1 & $^1P_1$       & P \\
SO           &       & 3 & $^1F_3$       & F \\
             &       & 5 & $^1H_5$       & H \\
             &       & 7 & $^1J_7$       & J \\
\hline
Singlet Even & (0,1) & 0 & $^1S_0$       & S \\
SE           &       & 2 & $^1D_2$       & D \\
             &       & 4 & $^1G_4$       & G \\
             &       & 6 & $^1I_6$       & I \\
\hline
Triplet Even & (1,0) & 0 & $^3S_1$       & S\\
TE           &       & 2 & $^3D_{1,2,3}$ & D \\
             &       & 4 & $^3G_{3,4,5}$ & G \\
             &       & 6 & $^3I_{5,6,7}$ & I \\
\hline
Triplet Odd  & (1,1) & 1 & $^3P_{0,1,2}$ & P \\
TO           &       & 3 & $^3F_{2,3,4}$ & F \\
             &       & 5 & $^3H_{4,5,6}$ & H \\
             &       & 7 & $^3J_{6,7,8}$ & J \\
\end{tabular}
\label{t:pwst}
\end{table}

\begin{figure*}[t]
\centering
\vspace{-5mm}
\includegraphics[width=.95\textwidth]{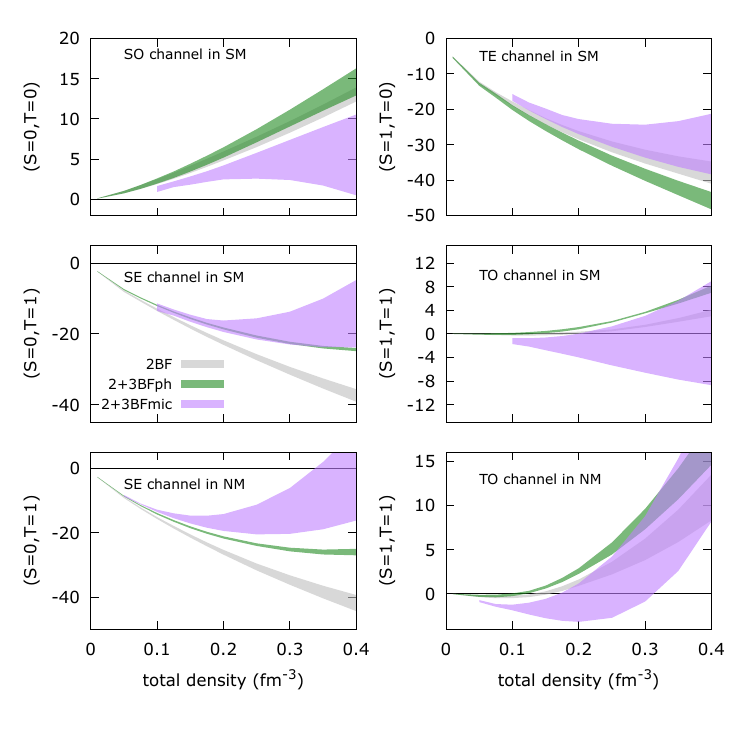}
\vspace{-10mm}
\caption{
Spin-isospin ($S,T$) channel decomposition
of the correlation energy per particle in MeV for SM and NM
as a function of the density and for the various interactions considered.
The bands quantify the internal accuracy of the BHF calculation
associated with the different 2BFs and 3BFs used.
}
\label{f:epot}
\end{figure*}

\subsection{Correlation energy per particle}

In the BHF approach the correlation energy per particle $\epot$,
last term of Eq.~(\ref{e:ebhf1}),
can be easily linked to the spin-isospin and the partial-wave decomposition
of the $G$-matrix,
providing an interesting insight into the impact
of the different $(S,T)$ channels and partial waves
contributing to the nuclear EOS.
With the aim of providing more stringent constraints for phenomenological EDFs,
we analyze in this section these decompositions for both SM and NM.
A particular application of these constraints to a Skyrme force
is discussed in the next section.

\subsubsection{Spin-isospin decomposition of the correlation energy}

The spin-isospin decomposition of the correlation energy per particle reads
\be
 \epot = \sum_{S,T} \epot^{(S,T)} \:,
\ee
where the contribution of each $(S,T)$ channel,
\be
 \epot^{(S,T)} = \frac{1}{2n\Omega} \sum_{i,j} f_i f_j\,
 \text{Re} \langle i j \vert G P_S P_T \vert i j \rangle_a \:,
\label{e:est}
\ee
can be obtained
by using the projectors on spin and isospin singlet and triplet states,
$P_S = \frac12 [ 1 + (S-\frac12) (1+\bm\sigma_1\cdot\bm\sigma_2) ]$
and
$P_T = \frac12 [ 1 + (T-\frac12) (1+\bm\tau_1\cdot\bm\tau_2) ]$.
The partial waves contributing to the different channels
are listed in table~\ref{t:pwst}.

The density dependence of the spin-isospin $(S,T)$ decomposition of $\epot$
in SM and NM is shown in Fig.~\ref{f:epot},
where the bands correspond to the uncertainty associated with
the different models used in our BHF calculation, as previously discussed.
The dotted and dashed lines in the figure refer to fits of Skyrme interactions,
which will be discussed in Sec.~\ref{sec:skyrme}.
In SM, the $(0,0)$ channel is repulsive
while the $(1,0)$ and $(0,1)$ channels are attractive.
The $(1,1)$ channel is much weaker than the others
and could be attractive or repulsive depending on the 3BF interactions
(it is slightly repulsive with 2BF and 2+3BFph).
The 2+3BFmic approach has a density dependence strong enough
to change the attractive $(0,1)$ channel for 2BF into a repulsive one
at high density,
or to turn the repulsive $(1,1)$ channel into an attractive one around $n_\sat$.
The $(0,1)$ and $(1,1)$ channels of NM are qualitatively similar
to those of SM.
The impact of the 3BF prescription can thus be opposite to the 2BF
in the various channels.
These features can be traced back to the general characteristics of the 3BF:
the two-pion-exchange 3BF component is attractive in the $(1,0)$
and repulsive in the $(0,1)$ channel,
and scalar repulsive 3BF components become dominant at high density
in all channels.

It is also interesting to note that the 2+3BFmic approach
has much wider error bars in the $(S,T)$ decomposition of $\epot$
than in the total $\epot$ shown in Fig.~\ref{f:ba}
due to strong compensation between the different channels,
which makes the total binding energy less dispersed
than the $(S,T)$ components.

\subsubsection{Partial-wave decomposition of the correlation energy}

\begin{figure}[t]
\centering
\vspace{-3mm}
\includegraphics[width=.95\textwidth]{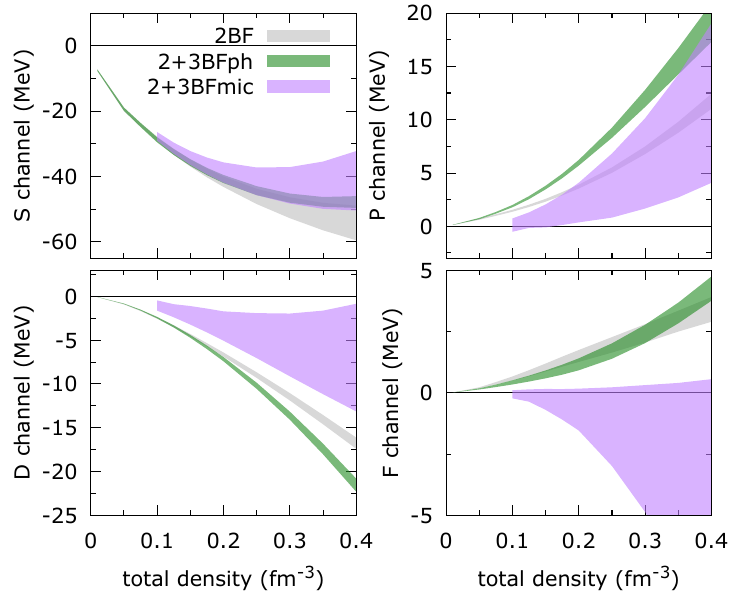}
\vspace{-1mm}
\caption{
Decomposition in fixed-$L$ partial waves
of the correlation energy per particle
in SM as a function of the density for the various interactions.
}
\label{f:epwL}
\end{figure}

\begin{figure}[t]
\centering
\vspace{-3mm}
\includegraphics[width=.95\textwidth]{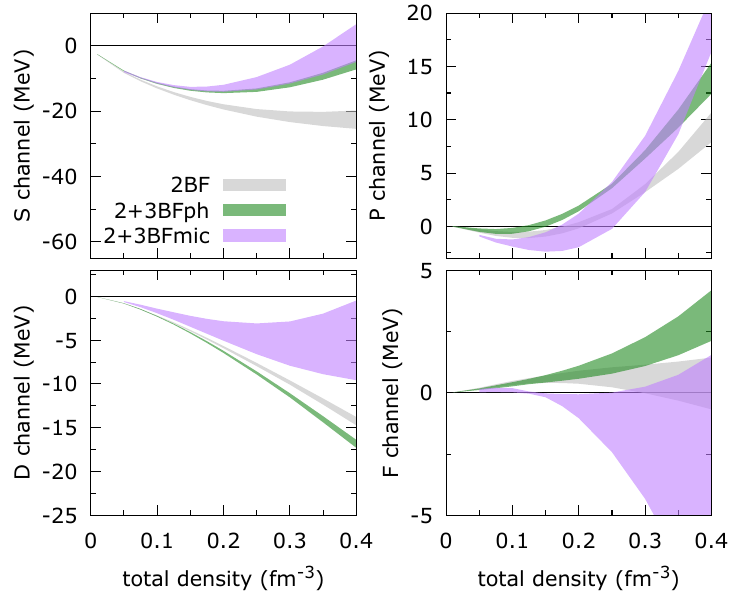}
\vspace{-1mm}
\caption{
Same as Fig.~\ref{f:epwL} in NM.}
\label{f:epwnL}
\end{figure}

Let us now further continue this analysis by looking in more detail
at the partial-wave decomposition of $\epot$ for both SM and NM.
The different partial waves contribute to the correlation energy as
\be
 \epot = \sum_{LSJ} \epot^{(LSJ)} \:,
\ee
where
\be
 \epot^{(LSJ)} = \frac{(2J+1)}{4n(2\pi)^6} \sum_{\tau\tau'}
 \int\!\!d^3\kv_\tau \, f_\tau(\kv_\tau)
 \int\!\!d^3\kv_{\tau'} \, f_{\tau'}(\kv_{\tau'})
 f_{\tau\tau'} \langle KqLSJ |
 G_{\tau\tau'\rightarrow\tau\tau'}
 | KqLSJ \rangle
\label{e:elj}
\ee
with $K=|\kv_\tau+\kv_{\tau'}|$,
$q =|(\kv_\tau - \kv_{\tau'})|/2$ and
$f_{\tau\tau'} = 1 + (-1)^{L+S} \delta_{\tau\tau'}$.

We first show in Figs.~\ref{f:epwL} (for SM) and \ref{f:epwnL} (for NM)
the decomposition into partial waves with fixed angular momentum $L$,
$\epot^{(L)} \equiv \sum_{SJ} \epot^{(LSJ)}$, i.e.,
all partial waves with the same value of $L$ are summed together.
Contributions up to a total angular momentum $J=8$ are considered
in the calculation,
although only results for the $S, P, D$, and $F$ partial waves are shown,
since the contributions of higher partial waves are small.
It is indeed clear from the figures that the contributions
of the different $\epot^{(L)}$ decrease as $L$ increases
for the range of densities considered here.
Note that the $P$ and $D$ channels are found to be of the same
order of magnitude in absolute value.
Since the standard Skyrme EDF contributes only to the $S$ and $P$ channels,
the relatively large importance of the $D$ channel
advocates for the extension of the standard Skyrme EDF
as suggested in Ref.~\cite{Davesne2015}.
The $F$ channel is found to be repulsive and less important than the others
for 2BF and 2+3BFph,
but strongly attractive and comparable to the $D$ channel
for 2+3BFmic.
This corresponds to the $(0,0)$ and $(1,1)$ channels in Fig.~\ref{f:epot}.

The most significant difference between SM and NM is in the $S$ channel,
which is less attractive in NM compared to SM,
due to the fact that the dominant attractive $^3S_1$ channel
contributes only to SM.
The other channels are qualitatively similar in SM and NM.
At low density $n<n_\sat$ one can clearly observe a dominance
of the $L=0$ contribution compared to the $L>0$ terms,
but as the density increases,
those terms become increasingly important.
As for the $(S,T)$ decomposition,
it is observed that the dispersion between the different predictions
of the partial waves is generally larger in SM compared to NM.

\begin{figure}[t]
\centering
\vspace{-3mm}
\includegraphics[width=.95\textwidth]{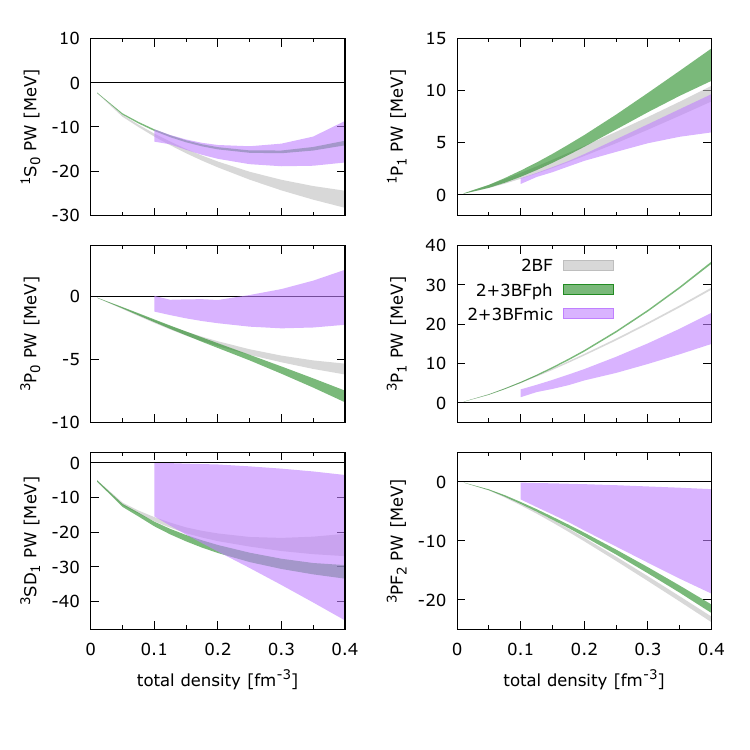}
\vspace{-10mm}
\caption{
Contributions to the correlation energy per particle from
various
partial waves,
as a function of the density for the different interactions.
}
\label{f:epw}
\end{figure}

Finally we present individual partial wave contributions to $\epot$
in Fig.~\ref{f:epw},
for the $^1S_0$, $^1P_1$, $^3P_0$, $^3P_1$, $^3SD_1$ and $^3PF_2$ waves.
It is only in the $^1S_0$ channel that the 3BF are similar,
while in the other channels they act in opposite way compared to the 2BF results.
The coupled $^3SD_1$ and $^3PF_2$ waves are predicted to be attractive
in all cases,
caused by the tensor force.
Note that the standard Skyrme EDFs contribute only to the $S$ and $P$ channels,
while in Ref.~\cite{Davesne2015}
the partial waves $^3P_0$ and $^3P_1$ have been used to calibrate
the spin-orbit coupling $W_0$ and the tensor parameter $t_o^{(n)}$
of extended EDFs.

\subsection{BHF single-particle energy}

\begin{figure}[t]
\centering
\vspace{-1mm}
\includegraphics[width=.95\textwidth]{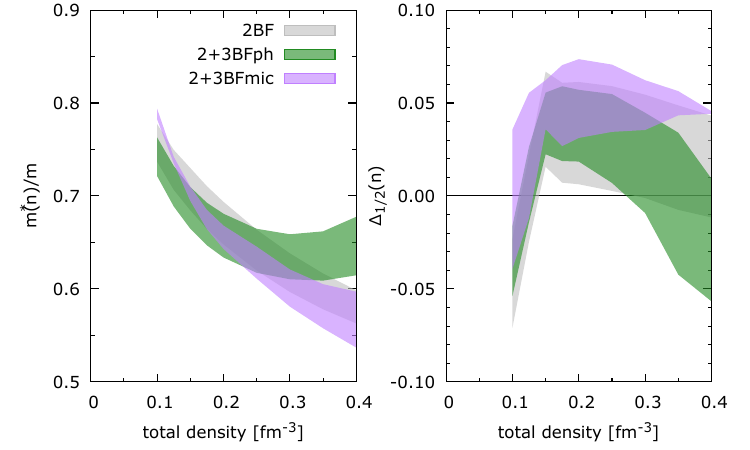}
\vspace{-1mm}
\caption{
(left) Effective mass in SM and
(right) isospin splitting of the effective mass for $\beta=0.5$
as a function of density for the various forces.
}
\label{f:figms}
\end{figure}

In the BHF approach,
the s.p.~energy $\eps_\tau(k)$ is defined by Eq.~(\ref{e:spe}).
In some phenomenological models however, like the Skyrme one,
the momentum dependence of $\eps_\tau$ is simply quadratic
and can be entirely incorporated in a modification of the mass
appearing in the kinetic term,
\be
 \eps_\tau(k) = \frac{k^2}{2m^*_\tau} + u_\tau \:,
\label{e:spep}
\ee
where the effective mass $m^*_\tau$ is a momentum-independent quantity
and the s.p.~potential $u_\tau$ is the value of the s.p.~potential at $k=0$.
Since we want to establish a link between microscopical and phenomenological
approaches,
using the BHF mean field potential $U_\tau(k)$ one can define
a global effective mass:
\bal
 u_\tau &= \mbox{Re}[U_\tau(k=0)] \:,
\\
 \frac{m^*_\tau}{m} &= \Big[ 1 + \frac{2m}{k_{F_\tau}^2}
 \mbox{Re}[U_\tau(k_{F_\tau}) - U_\tau(0)] \Big]^{-1} \:.
\label{e:effm2}
\eal
Note that a local momentum-independent effective mass
can also be determined from Eq.~(\ref{e:effm1})
as $m_\tau^* = m^*_\tau(k=k_{F_{\tau}})$.
This latter definition provides effective masses which are typically
5-10\% higher than using Eq.~(\ref{e:effm2}) at saturation density,
due to the presence of a wiggle in the BHF s.p.~potential close to $k_F$
\cite{wiggle}.
For the comparison with phenomenologial models,
we use the definition~(\ref{e:effm2})
for the global effective mass in the following.

The dependence of the effective mass on the isospin asymmetry
is also an interesting result to extract from the BHF approach.
One usually defines the effective-mass isospin splitting as
\be
 \Delta_\beta(n) \equiv \frac{m_n^*(n,\beta)-m_p^*(n,\beta)}{m} \:,
\ee
where $n=n_n+n_p$ is the total density and
$\beta=(n_n-n_p)/n$ is the isospin asymmetry parameter.
$\Delta_\beta(n)$ is often defined for maximal asymmetry $\beta=1$ (NM)
and at saturation density $n=n_\sat$.
Since this choice requires some technicalities for the definition
of the proton effective mass in a pure neutron environment,
we have preferred in this work to employ $\Delta_{1/2}(n_\sat)$,
using the  approximation
$\Delta_{1} \approx 2\Delta_{1/2}$ \cite{Zuo05}.
Note that this relation is exact for Skyrme-type interactions,
see Sec.~\ref{sec:skyrme}.

The density dependence of the effective mass in symmetric matter $m^*(n)/m$
and of the isospin splitting $\Delta_{1/2}(n)$
are shown in Fig.~\ref{f:figms}, panels (a) and (b), respectively.
At normal density the impact of 3BFs is small
for both the effective mass and the isospin splitting.
The effective mass is already well determined by 2BF only,
that predict a continuous decrease of the effective mass
and a positive sign for $\Delta_{1/2}(n)\approx 0.03\pm0.03$ above $n_\sat$.
While the differences between 2+3BFph and 2+3BFmic predictions
are small below saturation density,
at larger density
the 2+3BFmic results remain close to the 2BF ones,
while 2+3BFph predict a rise of the effective mass
and a change of sign of $\Delta_{1/2}$ above $2n_\sat$.

We therefore conclude that the momentum dependence of the 3BFph results
is more marked than the one for 3BFmic.
In the next subsection,
we show that this is due to a compensation between two $(S,T)$ contributions
for 3BFmic,
which is absent in the case of 3BFph.
As a consequence,
as the density increases
3BFph tends to reduce the correction due to the effective mass
and the difference between the momentum-dependent contributions
from neutrons and protons in asymmetric matter.

\subsubsection{Spin-isospin decomposition of the effective mass}

\begin{figure}[t]
\centering
\vspace{-5mm}
\includegraphics[width=.95\textwidth]{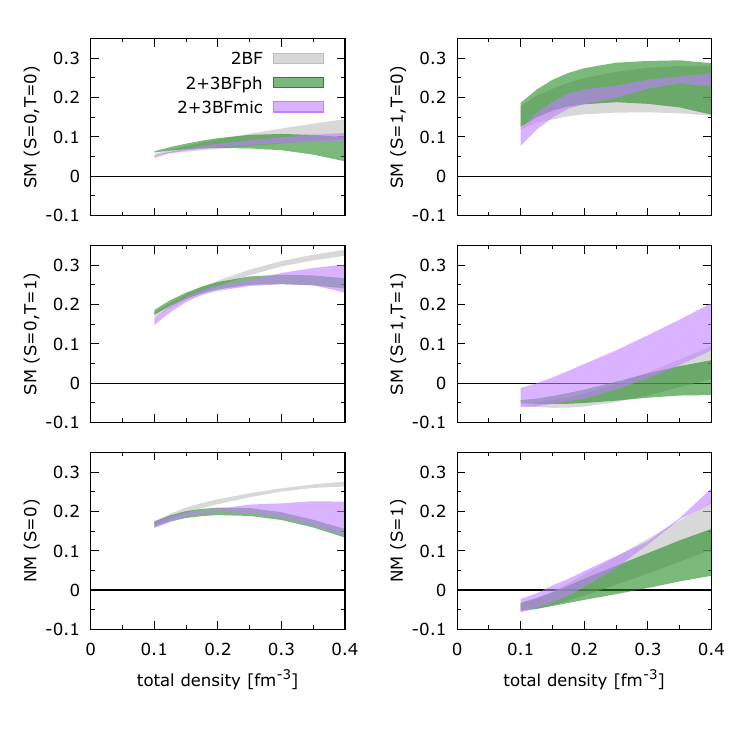}
\vspace{-10mm}
\caption{
Spin-isospin decomposition of the effective mass $(m/m^*)^{(S,T)}$,
Eq.~(\ref{e:mst}), in SM and NM
as a function of the density
for various interactions.
}
\label{f:effmass}
\end{figure}

In Fig.~\ref{f:effmass} we show the spin-isospin decomposition
of the inverse of the effective mass
\be
 \frac{m}{m^*_\tau} = 1 + \sum_{S,T}
 \left(\frac{m}{m^*_\tau}\right)^{(S,T)} \:,
\ee
where
\be
 \left(\frac{m}{m^*_\tau}\right)^{(S,T)} \equiv
 \frac{2m}{k_{F_\tau}^2}
 \text{Re} \big[ U^{(S,T)}_\tau(k_{F_\tau}) - U^{(S,T)}_\tau(0) \big] \:.
\label{e:mst}
\ee
being $U^{(S,T)}_\tau(k)$ the contribution of the spin-isospin channel $(S,T)$
to the s.p.~mean field.

In the considered density range the inverse of the effective mass is dominated
by the $S$-wave contributions in the $(0,1)$ and $(1,0)$ channels,
where there is a reasonable agreement between the two 3BF prescriptions.
Note that in the SM $(0,1)$ channel the two 3BF predict a increase
of the inverse of the effective mass compared to the 2BF above $n_\sat$:
this originates from an attractive term in the 3BF,
which moderates the repulsive contribution of the 2BF.

In the SM $(1,1)$ (and NM $(S=1)$) channels the 2+3BFmic
has a stronger density dependence than 2+3BFph.
The larger contribution to $m/m^*$ indicates
that there is an additional repulsive momentum dependence in 3BFmic
that is absent from 3BFph.
For 3BFmic, it happens that this additional repulsive contribution
in the $(1,1)$ channel compensates the additional attraction in $(0,1)$,
such that the sum of all contributions to $m/m^*$
is similar for 2BF and 2+3BFmic,
as remarked in the discussion of Fig.~\ref{f:figms}.
For 3BFph, the absence of the repulsive contribution in the $(1,1)$ channel
induces the marked difference between 2BF and 2+3BFph
for the inverse of the effective mass
shown in Fig.~\ref{f:figms}.
In the following, we show that the more repulsive 3BFmic
leads also to more repulsion for the total energy per particle.

\subsection{Total energy}

\begin{figure}[t]
\centering
\vspace{-1mm}
\includegraphics[width=.95\textwidth]{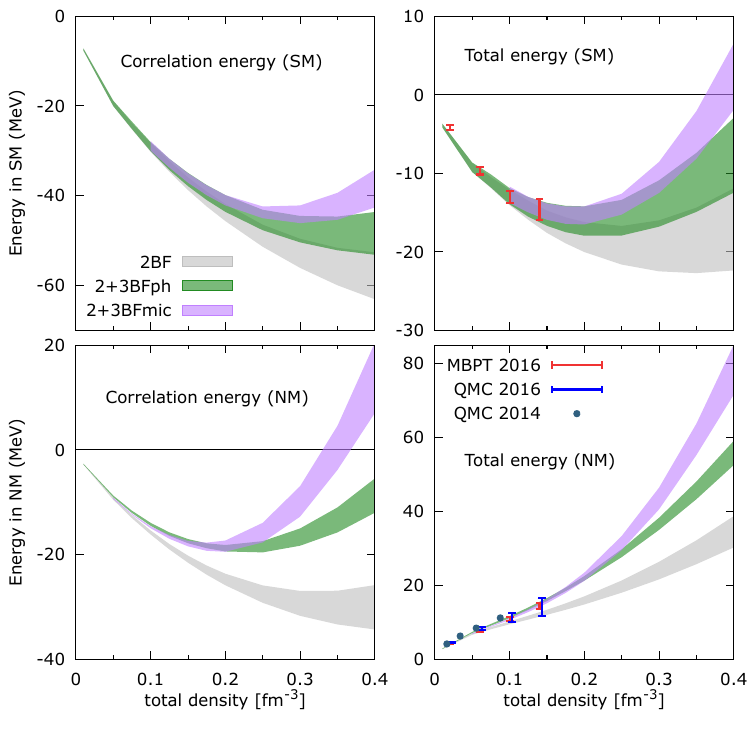}
\vspace{-5mm}
\caption{
BHF correlation energy (left panels)
and total energy (right panels)
in SM (upper panels) and NM (lower panels)
as a function of the density for the various $NN$ potentials only (2BF)
and including 3BF (2+3BFph, 2+3BFmic).
Several $\chi$-EFT results
\cite{Drischler2016,Tews2016,Wlazlowski2016}
are also shown for comparison,
see text for more details.
}
\label{f:ba}
\end{figure}

We show in Fig.~\ref{f:ba} the correlation and total energy per particle
of SM and NM as function of the density
for the various 2BFs with and without the contribution of 3BFs.
The SM results with only 2BF exhibit a saturation density
well above the empirical value $n_\sat$.
This is well known and due to the different off-shell behavior of the 2BF:
phase-shift-equivalent potentials lead to different saturation curves
with the corresponding saturation points lying in the so-called ``Coester band"
\cite{li06}.
The band defined by these curves can be used to estimate the contribution of 2BF
to the uncertainty in the energy per particle.
As expected the error bar is small at low densities,
where the interaction is mainly dominated by the $S$ partial waves,
but increases rapidly with density,
where higher partial waves become important.
Similar behavior is observed for the 3BFph and 3BFmic results,
and also for NM,
where the uncertainty bands are narrower due to the weaker interaction.
Above $n_\sat$ we observe that 2+3BFmic predict more repulsion than 2+3BFph,
as discussed in the previous subsection.

At low density, our predictions are compared to $\chi$-EFT predictions
based on 2+3BF calculated within different orders and approaches:
\begin{itemize}
\item (a) N2LO Wlaz\l{}owski QMC 2014 \cite{Wlazlowski2016}:
Variational QMC calculation of NM
with chiral nuclear forces at N2LO for 2BF and 3BF.
\item (b) N2LO Tews QMC 2016 \cite{Tews2016}:
AFDMC calculation of NM
with chiral nuclear forces at N2LO for 2BF.
The 3BF is taken at leading order and local.
\item (c) N3LO Drischler MBPT 2016 \cite{Drischler2016}:
Many-body perturbation theory approach based on regularized
chiral nuclear forces at N3LO for SM and NM.
\end{itemize}

Note that while there are no uncertainties estimated in (a),
the estimation of the uncertainty is performed differently
in (b) and (c) compared to our BHF results.
In those cases,
the uncertainty includes the unknown from the chiral 2BF fitted on phase shifts
as well as the uncertainty in the truncation order,
which is expected to decrease as the order increases.
Thus it is not surprising that the N3LO calculation \cite{Drischler2016}
has slightly lower error bars than the N2LO one \cite{Tews2016}.
There are also differences in the way the band widths are estimated
among these calculations.
It is however interesting to note that the band of uncertainty
in our case is compatible with the most advanced
N3LO MBPT calculation \cite{Drischler2016} in both SM and NM.
This provides a solid ground for our extrapolation to higher density,
even if one cannot exclude an additional uncertainty
related to the theoretical approach itself.

The conclusion of this part is that there are differences
between the 3BF considered in this paper, 3BFph and 3BFmic,
due to their different construction.
The 3BFmic predict a more repulsive energy per particle than the 3BFph
at high density.

\section{Constraining the Skyrme energy density functional
from BHF calculations}
\label{sec:skyrme}

The commonly called standard Skyrme interaction is a contact interaction
with momentum-dependent terms,
which is expressed in the following form,
\bal
 V(\rv_1,\rv_2) &=
 t_0(1+x_0P_\sigma)\; \delta(\rv)
\nonumber \\
 &+ \frac{t_1}{2}(1+x_1P_\sigma)\; \big[\kv'^2\delta(\rv)+\delta(\rv)\kv^2\big]
\nonumber \\
 &+ t_2(1+x_2P_\sigma)\; \kv'\cdot \delta(\rv)\kv
\nonumber\\
 &+ \frac{t_3}{6}(1+x_3P_\sigma)\; n(\Rv)^\gamma \delta(\rv)
\nonumber \\
 &+ iW_0(\sv_1+\sv_2)\cdot[\kv'\times\delta(\rv)\kv] \ ,
\label{e:skm}
\eal
where
$\rv=\rv_1-\rv_2$, $\Rv=(\rv_1+\rv_2)/2$,
$\kv=(\bm\nabla_1-\bm\nabla_2)/i$ is the relative momentum acting on the right,
$\kv'$ its conjugate acting on the left,
and $P_\sigma=(1+\sv_1\cdot\sv_2)/2$ is the spin-exchange operator.
The last term, proportional to $W_0$,
corresponds to the zero-range spin-orbit term.
It does not contribute to the EOS in homogeneous systems
and thus will be ignored for the rest of this article.
In its standard form \eqref{e:skm} the Skyrme interaction contributes only
in the $L=0$ and $L=1$ channels,
which implies some limitations which will appear in the following discussion.

In the Hartree-Fock approximation,
the total energy density of SM ($\beta=0$) and NM ($\beta=1$)
is given respectively as
\bal
 \eps(n,\beta=0) &=
 \frac{1}{2m}\tausm + C_0^\tau n\tausm + C_0^n(n) n^2  \:,
\label{e:easksm}
\\
 \eps(n,\beta=1) &=
 \frac{1}{2m}\taunm + [C_0^\tau + C_1^\tau] n\taunm
 + \big[C_0^n(n)+C_1^{n}(n)\big]n^2  \:,
\label{e:easknm}
\eal
where the kinetic densities are
$\tausm=\frac{3}{5}(\frac{3\pi^2}{2})^{2/3}n^{5/3}$ and
$\taunm=\frac{3}{5}(3\pi^2)^{2/3}n^{5/3}$, and
$C_t^n(n)=C_{t0}^n+C_{t3}^n n^\gamma$ for $t=0,1$.
In this work, we employ the DFT coefficients,
e.g. $C_t^\tau$, $C_{t0}^n$, and $C_{t3}^n$,
which can be defined in terms of the Skyrme parameters $t_i$ and $x_i$,
see Ref.~\cite{Bender2003}, where they have been introduced.

Several authors
(see, e.g., Refs.~\cite{Cao2006,Lesinski2006,Goriely2010,Gambacurta2011})
have stressed the interest of using the spin-isospin decomposition
of the BHF correlation energy as an additional constraint
to better determine the parameters of energy density functionals
based on effective interactions such as the Skyrme one.
It indeed allows a more detailed determination of the parameters
entering into the calculation of the energy per particle
by increasing the number of equations to solve.
From Eq.~(\ref{e:est}),
one can calculate the decomposition of the correlation energy density in terms of
SE $\epspot^{(0,1)}$,
TE $\epspot^{(1,0)}$,
SO $\epspot^{(0,0)}$, and
TO $\epspot^{(1,1)}$ contributions as \cite{Lesinski2006}
\bal
 \epspot^{(0,0)} &=
 \frac{1}{16} \left[ C_0^\tau-3C_0^{sT}-3C_1^\tau+9C_1^{sT}\right] n\tausm \:,
\nonumber\\
 \epspot^{(0,1)} &=
 \frac{3}{4} \left[ C_0^n(n)+C_1^{n}(n) \right] n^2
+ \frac{3}{16} \left[ C_0^\tau-3C_0^{sT}+C_1^\tau-3C_1^{sT}\right] n\tausm \:,
\nonumber\\
 \epspot^{(1,0)} &=
 \frac{1}{4} \left[ C_0^n(n) - 3C_1^{n}(n) \right] n^2
 + \frac{3}{16} \left[ C_0^\tau+C_0^{sT}-3C_1^\tau-3C_1^{sT}\right] n\tausm \:,
\nonumber \\
 \epspot^{(1,1)} &=
 \frac{9}{16} \left[ C_0^\tau+C_0^{sT}+C_1^\tau+C_1^{sT}\right] n\tausm \:
\label{e:esm11}
\eal
in SM, and as
\bal
 \epspot^{(0)} &=
 \left[ C_0^n(n)+C_1^n(n) \right] n^2
+ \frac{1}{4} \left[ C_0^\tau-3C_0^{sT}+C_1^\tau-3C_1^{sT}\right] n\taunm
\nonumber\\
 \epspot^{(1)} &=
 \frac{3}{4} \left[ C_0^\tau+C_0^{sT}+C_1^\tau+C_1^{sT}\right] n\taunm
\label{e:enm11}
\eal
in NM.
The correlation energy in the channels $(0,0)$ and $(1,1)$ in SM,
and $(1)$ in NM scales only with the kinetic energy, $\tausm$ and $\taunm$,
because only the angular momentum $L=1$ can contribute in these channels.
In the channels $(0,1)$ and $(1,0)$ in SM, and $(0)$ in NM,
there is also a term that scales with the density $n$
reflecting the $L=0$ contribution of the Skyrme interaction.
Depending on the sign of the combination of the coefficients,
the energy per particle in the $(S,T)$ channels can be positive or negative,
in the channels $(0,0)$ and $(1,1)$ in SM, and $(1)$ in NM,
or can change its sign as a function of the density in the other channels.
This is a clear limitation of the standard Skyrme interaction \eqref{e:skm},
since BHF calculations predict a richer density dependence,
see Fig.~\ref{f:epot} for instance.

In nuclear matter,
the Skyrme interaction is determined by 8 coefficients 
$C_t^\tau$, $C_t^{sT}$, $C_{t0}^\rho$, and $C_{t3}^\rho$,
as well as the exponent of the density dependence $\gamma$.
We have 6 constraints in terms of the $(S,T)$ decomposition
and we decide to fix the value of the parameter $\gamma$, see hereafter.
We therefore need two additional constraints to determine unambiguously
all Skyrme parameters.
These two other constraints will be given by the effective mass in SM
and the isospin splitting,
namely from the BHF predictions we fix
$m^*_\sat/m=0.7$ and $\Delta_{1,\sat}=0.1$ for 2BF, 2+3BFph, and 2+3BFmic.

For the Skyrme interaction,
the nucleon effective mass is expressed as \cite{Bender2003,Lesinski2006}
\be
 \frac{1}{2m_\tau^*(n,\beta)} = \frac{1}{2m} +
 \left[ C_0^\tau + \tau_3 C_1^\tau \beta \right] n \:.
\ee
The parameters $C_0^\tau$ and $C_1^\tau$ can therefore be obtained as
\bal
 C_0^\tau &= \frac{1}{n_\sat} \frac{1}{2m}
 \left[ \frac{m}{m^*_\sat} - 1 \right] \:,
\label{e:c0tau}
\\
 C_1^\tau &= \frac{1}{n_\sat}\frac{1}{2m}\frac{1}{\beta \Delta_\beta}
 \left[ 1 - \sqrt{1 + \left(\frac{\Delta_\beta}{m^*_\sat/m}\right)^2} \right]
\\
 &\approx -\frac{1}{n_\sat}\frac{1}{2m}\frac{\Delta_\beta}{2\beta}
 \left( \frac{m}{m_\sat^*} \right)^2  \:,
\label{e:c1tau}
\eal
where the last approximation is valid for typical small values of
$\left(\frac{\Delta_\beta}{m^*_\sat/m}\right)^2 \approx 0.02$.

\begin{figure}[t]
\centering
\includegraphics[width=.95\textwidth]{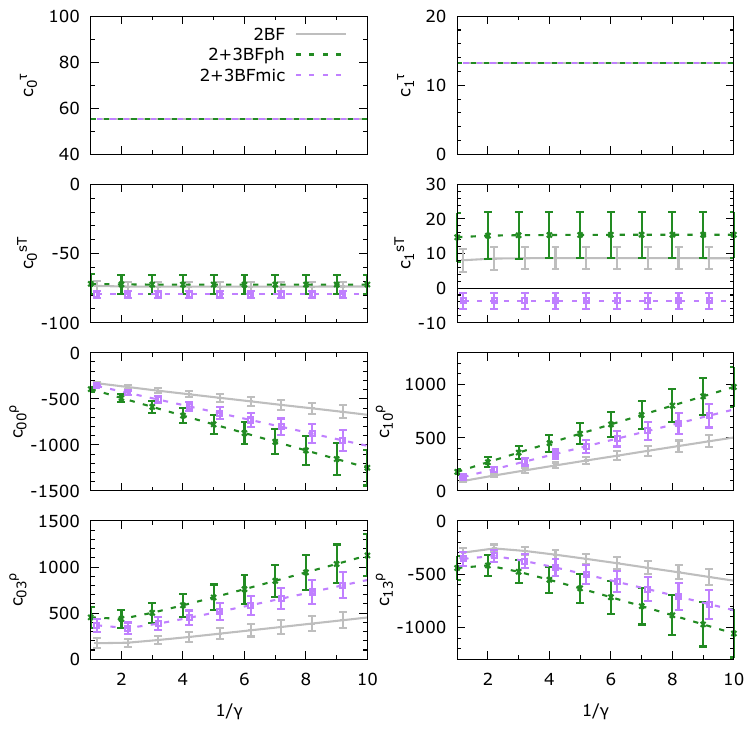}
\vspace{-7mm}
\caption{
Skyrme coefficients obtained from the fit of $\epspot^{(S,T)}$
and the effective mass as function of $1/\gamma$.
The models are calibrated to the following quantities;
$m^*_\sat/m=0.7$ and $\Delta_{1,\sat}=0.1$,
and to the BHF correlation energies in the $(S,T)$ channels
in SM and $(S)$ channels in NM.}
\label{f:fit}
\end{figure}

We determine the 8 Skyrme coefficients by imposing the global reproduction
of the 6 $(S,T)$ channels predicted by the BHF calculations
in the density range going from 0.01 up to 0.4~fm$^{-3}$ in SM and NM
(including uncertainties),
complemented by the effective mass $m^*_\sat/m$
and its isospin splitting $\Delta_{1,\sat}$.
The fit is provided by the nonlinear least-squares Marquardt-Levenberg algorithm
encoded in Gnuplot
\footnote{see https://gnuplot.sourceforge.net/docs\_4.2/node82.html}.
The results are shown in Fig.~\ref{f:fit},
where the parameter $\gamma$ is explored from $1$ to $1/10$.
The error bars are estimated from the covariant matrix properties
at the solution.
There are 4 Skyrme coefficients $C_t^\tau$ and $C_t^{sT}$,
with $t=0,1$, which are independent of $\gamma$,
and 4 other coefficients $C_{t0}^\rho$ and $C_{t3}^\rho$,
which depend on the value taken for $\gamma$.
We also observe a strong impact of the 3BF for the coefficient $C_1^{sT}$,
and a noticeable but less strong impact for the coefficients
$C_{t0}^\rho$ and $C_{t3}^\rho$.

\begin{figure}[t]
\centering
\includegraphics[width=.95\textwidth]{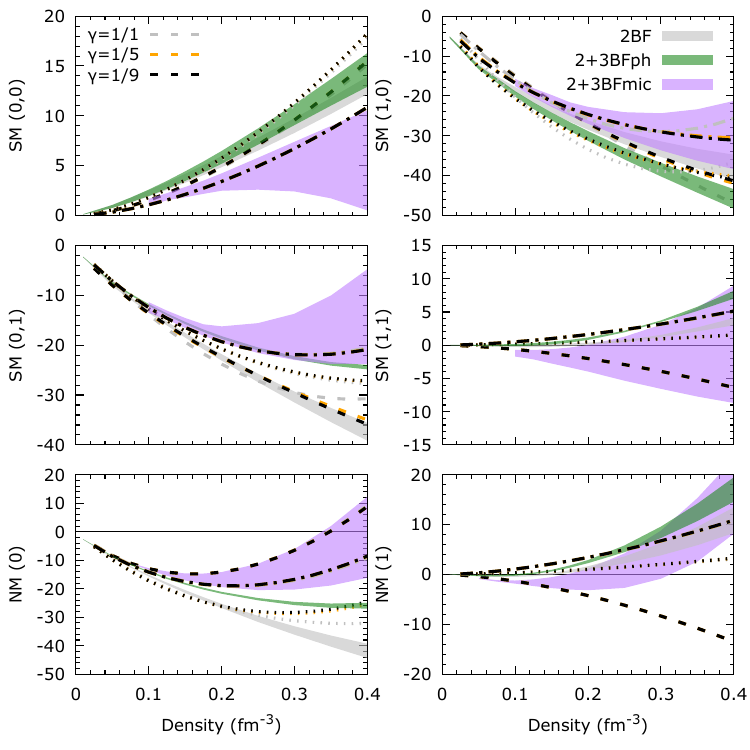}
\vspace{-3mm}
\caption{
Spin-isospin ($S,T$) channel decomposition
of the correlation energy per particle in SM and NM as a function of the density
for the various interactions considered.
The bands quantify the internal accuracy of the BHF calculation
associated with the different 2BFs and 3BFs used.
The curves are the result of the fit of the Skyrme EDF:
the dashed lines represent fits of 2BF, dotted of 2+3BFph,
and dashed-dotted of 2+3BFmic.}
\label{f:fitepot}
\end{figure}

In order to evaluate the quality of the fits,
we compare in Fig.~\ref{f:fitepot} the quantities targeted by our fits
and the results provided by our fits.
The Skyrme spin-isospin $(S,T)$ channel decomposition
of the correlation energy per particle in SM and NM are shown
together with the constraints extracted from BHF calculations.
We consider the centroid of the Skyrme coefficients
obtained from Fig.~\ref{f:fit} and vary the parameter $\gamma$.
The effect of $\gamma$ is very weak since it is absorbed in the values of the coefficients $C_{t0}^n$ and $C_{t3}^n$, see Fig.~\ref{f:fit}.
The other four coefficients $C_t^\tau$ and $C_t^{sT}$ are independent of $\gamma$.
The quality of the fit is however not very good.
This is particularly clear in the following channels:
$(0,0)$, $(1,1)$ in SM and $(1)$ in NM.
These channels are the ones where only the $L=1$ contribution
of the Skyrme interaction plays a role.
The poor reproduction of the BHF results is therefore
due to the lack of flexibility of the standard form \eqref{e:skm}
of the Skyrme interaction, where higher values of $L$ are necessary.

Let us remark from Fig.~\ref{f:fitepot} that the 2BF is quite well reproduced
by the Skyrme model in SM and for the $(0,0)$, $(0,1)$ and $(1,0)$ channels.
The $(1,1)$ channel has a wrong sign in the Skyrme model,
although the strength of the interaction in this channel is quite weak.
In NM, however, the fit of the 2BF is poor above $n_\sat$.
We, therefore, conclude that already at the 2BF level
the Skyrme interaction could not reproduce well the BHF predictions
in both SM and NM above $n_\sat$.

In Ref.~\cite{Goriely2010}, a set of Skyrme interactions have been obtained by employing the $(S,T)$ decomposition in SM for two BHF calculations and the NM channel was controlled globally by variational calculations. The difference with our approach are i) that our uncertainty estimate is based on a larger number of calculations, and ii) that we employ the $(S)$ channel decomposition in NM to adjust Skyrme EDF.
In Ref.~\cite{Gambacurta2011}, only one BHF calculation is employed to fit the Skyrme EDF and only the $(S,T)$ decomposition in SM is employed.
In Refs.~\cite{Goriely2010,Gambacurta2011} data from finite nuclei, e.g. binding energies and charge radii, have been employed in addition to the BHF constraints. It is therefore difficult to perform detailed comparisons of our results with the ones presented in these papers. We however conclude from our analysis that the standard Skyrme EDF should be enriched, with higher $L$ terms for instance, in order to reproduce BHF calculations in $(S,T)$ channel in SM and NM up to at least 0.4~fm$^{-3}$. The results presented in this paper can be employed for such aim.

\section{Conclusions}
\label{sec5}

In this paper, we have analyzed BHF results of SM and NM
by decomposing the total energy and the effective mass
into spin-isospin $(S,T)$ channels
and also into contributions from various partial waves.
The two-body and three-body contributions were shown separately
as bands representing the uncertainties after exploring a set of interactions.
We found that 3BFmic induces more repulsion than 3BFph,
visible in the energy per particle at high density.
This repulsion originates from a stronger repulsive scalar-meson-exchange
term of 3BFmic compared to 3BFph.
The effective mass, however, seems to be more impacted by 3BFph than by 3BFmic.
We show that this results from a compensation between two channels
in 3BFmic which is absent for 3BFph.

We have then addressed the question of the effectiveness
of the standard Skyrme model \eqref{e:skm}
to reproduce the $(S,T)$ decompositions
predicted by BHF calculations based on different 2BFs and 3BFs.
Our results indicate that the standard Skyrme model can not reproduce
the $(S,T)$ decomposition predicted by BHF calculations above $n_\sat$
considering only 2BF.
This conclusion is even stronger when 3BFs are added.

The detailed and numerous quantities obtained from BHF calculations
and presented in this paper can however serve as guidelines
for improvements of the Skyrme interaction with close links to BHF calculations.
Having reliable models to describe the properties of baryonic matter
at extreme densities on the one hand,
and fair estimates of the theoretical uncertainties on the other hand,
are important for the treatment of the astrophysical signals
that will be collected in the future.


\authorcontributions{
All authors contributed to this work.
All authors have read and agreed to the published version of the manuscript.
}

\funding{
JM is partially supported by CNRS-IN2P3 MAC masterproject,
as well as from the LABEX Lyon Institute of Origins (ANR-10-LABX-0066).
}

\acknowledgments{
This work was also fostered by the scientific emulation created by the
``NewCompStar" COST Action MP1304 and PHAROS COST Action MP16214.
}

\conflictsofinterest{The authors declare no conflicts of interest.}


\reftitle{References}


\end{document}